\documentclass[hyper]{JHEP3} 
\preprint{}
\JHEPspecialurl{http://jhep.sissa.it/JOURNAL/JHEP3.tar.gz}
\usepackage{epsfig,multicol,bbm}
\newcommand\fverb{\setbox\pippobox=\hbox\bgroup\verb}
\newcommand\fverbdo{\egroup\medskip\noindent%
\fbox{\unhbox\pippobox}\ }
\newcommand\fverbit{\egroup\item[\fbox{\unhbox\pippobox}]}
\newbox\pippobox

\title{Similarities Between Classical Timelike Geodesics in a Naked
Reissner-Nordstrom Singularity Background and the Behaviour of Electrons in Quantum Theory}

\author{Asghar Qadir$^{1,2}$ and Azad A. Siddiqui$^{3,4}$~\\
$^{1}$Center for Advanced Mathematics and Physics, Campus of EME
College, National University of Science and Technology, Peshawar
Road, Rawalpindi, Pakistan\\\\  $^{2}$Department of Mathematical
Sciences, King Fahd University of Petroleum and Minerals, Dhahran,
31261, Saudi Arabia\\\\
$^{3}$Department of Basic Science and Humanities, EME College,
National University
of Science and Technology, Peshawar Road, Rawalpindi, Pakistan\\\\
$^{4}$Department of Physics and Measurement Technology, Linköping
University, SE-583 81 Linköping, Sweden\\

{\tt aqadirs@comsats.net.pk, azad@ifm.liu.se, azad-eme@nust.edu.pk}}

\abstract{It is generally assumed that naked
singularities must be physically excluded, as they could otherwise
introduce unpredictable influences in their future null cones.
Considering geodesics for a naked Reissner-Nordstrom singularity,
it is found that the singularity is effectively clothed by its repulsive
nature. Regarding electron as naked singularity, the size of
the clothed singularity (electron) turns out to be classical
electro-magnetic radius of the electron, to an observer falling freely from infinity,
initially at rest. The size shrinks for an observer falling freely from infinity,
with a positive initial velocity. For geodetic parameters corresponding to negative energy
there are trapped geodesics. The similarity of this picture with
that arising in the Quantum Theory is discussed.}


\begin{document}

\section{Introduction}

Penrose proposed the cosmic censorship hypothesis \cite{Penrose1} so as to avoid
the possibility of unpredictable influences emerging from the
singularity, where physical laws break down. As he put it \cite{Penrose2}, ``it
is as if there is a cosmic censor board that objects to naked
singularities being seen and ensures that they only appear
suitably clothed by an event horizon". Due to this conjecture,
naked singularities are seldom studied seriously in themselves,
though various discussions focus on the possibility of finding
counter-examples to it even for singularities that arise from
realistic gravitational collapse processes \cite{Joshi}. The geodesics of arbitrarily charged particles
in a naked Reissner-Nordstrom singularity background are studied in \cite{Cohen}. However, it
concentrates on calculating the geodesics only and not on deducing
any consequences from them. While attempting to foliate the
Reissner-Nordstrom geometry by flat spacelike hypersurfaces \cite{Azad} for a
usual black hole we found it necessary to investigate geodesics in
a naked singularity background. Here we report on the striking
similarity of the behaviour of these geodesics and the Quantum
picture for an electron.

The fact that two such apparently different theories as General
Relativity (GR) and Quantum Theory (QT) come up with unexpectedly
similar features seems remarkable to us. In particular, GR dealing
with point particles seems to require (as will be shown
subsequently) that they acquire an extension and that their
interactions involve non-local effects. This is not to say that we
claim that QT derives from GR or vice versa. Rather, we note that
the similarity of the pictures suggest that some more fundamental
theory yields both. While the spatial extension is, in some sense
apparent for a ''clothed'' black hole it is not directly apparent
for the naked singularity. Since a GR description of an electron
would be as a naked singularity, this fact is of importance.

\section{Timelike Geodesics}

We start by giving a brief review of unforced timelike geodesics
(corresponding to the paths of uncharged test particles) in the
Reissner-Nordstrom background. The metric is taken in the usual
form
\begin{equation}
ds^{2}=e^{\upsilon \left( r\right) }dt^{2}-e^{-\upsilon \left(
r\right) }dr^{2}-r^{2}d\theta ^{2}-r^{2}\sin ^{2}\theta d\varphi
^{2},\label{1}
\end{equation}
where $e^{\upsilon \left( r\right) }=1-2m/r+Q^{2}/r^{2}$. The
geodesic equation for $t$ gives
\begin{equation}
\frac{dt}{ds}=\dot{t}=ke^{-\upsilon \left( r\right) }.  \label{6}
\end{equation}
For freely falling observers we take $k=1$ so as to obtain the
flat spacetime value of $\stackrel{.}{t}$ at infinity. Had there
been a finite velocity at infinity, $\stackrel{.}{t}$ would have been greater than
unity and consequently, we would have to take $k>1$.

From Eq.$\left( \ref{1}\right) $, for $\theta =\phi =$ constant, we have
\begin{equation}
e^{\upsilon \left( r\right) }\dot{t}^{2}-e^{-\upsilon \left( r\right) }\dot{r
}^{2}=1.  \label{7}
\end{equation}
We can re-write the two requirements, Eqs.($\ref{6}$ and
$\ref{7}$), in a single equation for the change of $r$ with $t$, as
\begin{equation}
\frac{dr}{dt}=\frac{\stackrel{.}{r}}{\stackrel{.}{t}}=\frac{\pm \sqrt{
k^{2}-e^{\upsilon \left( r\right) }}}{ke^{-\upsilon \left(
r\right) }}.  \label{8}
\end{equation}
It is clear that the geodesics will be defined only for
$e^{\upsilon \left( r\right) }<k^{2}$.\ We see that we must take
the negative root, on account of the initial attraction of the
gravitational source. If we take $k=1$,\ as required for observers
falling freely from infinity (initially at rest), then there is obviously a boundary
at
\begin{equation}
r_{b}=\frac{Q^{2}}{2m},
\end{equation}
at which $dr/dt$ becomes zero and after which the positive root
has to be chosen.

The geodesics can be better represented in terms of a re-scaled
radial parameter,
\begin{equation}
r^{\ast }=\int e^{-\upsilon \left( r\right) }dr=\int \left(
1-2m/r+Q^{2}/r^{2}\right) ^{-1}dr.
\end{equation}
Here, the constant of integration is chosen so that $r^{\ast }=0$ at $
r=0$. Therefore,

\begin{eqnarray}
r^{\ast } &=&r+m\ln \left( \frac{r^{2}-2mr+Q^{2}}{Q^{2}}\right) +\frac{
2m^{2}-Q^{2}}{\sqrt{Q^{2}-m^{2}}}  \nonumber \\
&&\times \left[ \tan ^{-1}\left(
\frac{r-m}{\sqrt{Q^{2}-m^{2}}}\right) -\tan ^{-1}\left(
\frac{m}{\sqrt{Q^{2}-m^{2}}}\right) \right].\nonumber
\\\label{9a}
\end{eqnarray}

Now, taking $k=1$, Eq.$\left(\ref{8}\right) $ can be re-written, in $\left( t,r^{\ast }\right) $ coordinates, as
\begin{equation}
\frac{dr^{\ast }}{dt}=\frac{\stackrel{.}{r}^{\ast
}}{\stackrel{.}{t}}=\pm \sqrt{\frac{2m}{r}-\frac{Q^{2}}{r^{2}}.}
\end{equation}
\DOUBLEFIGURE[t]{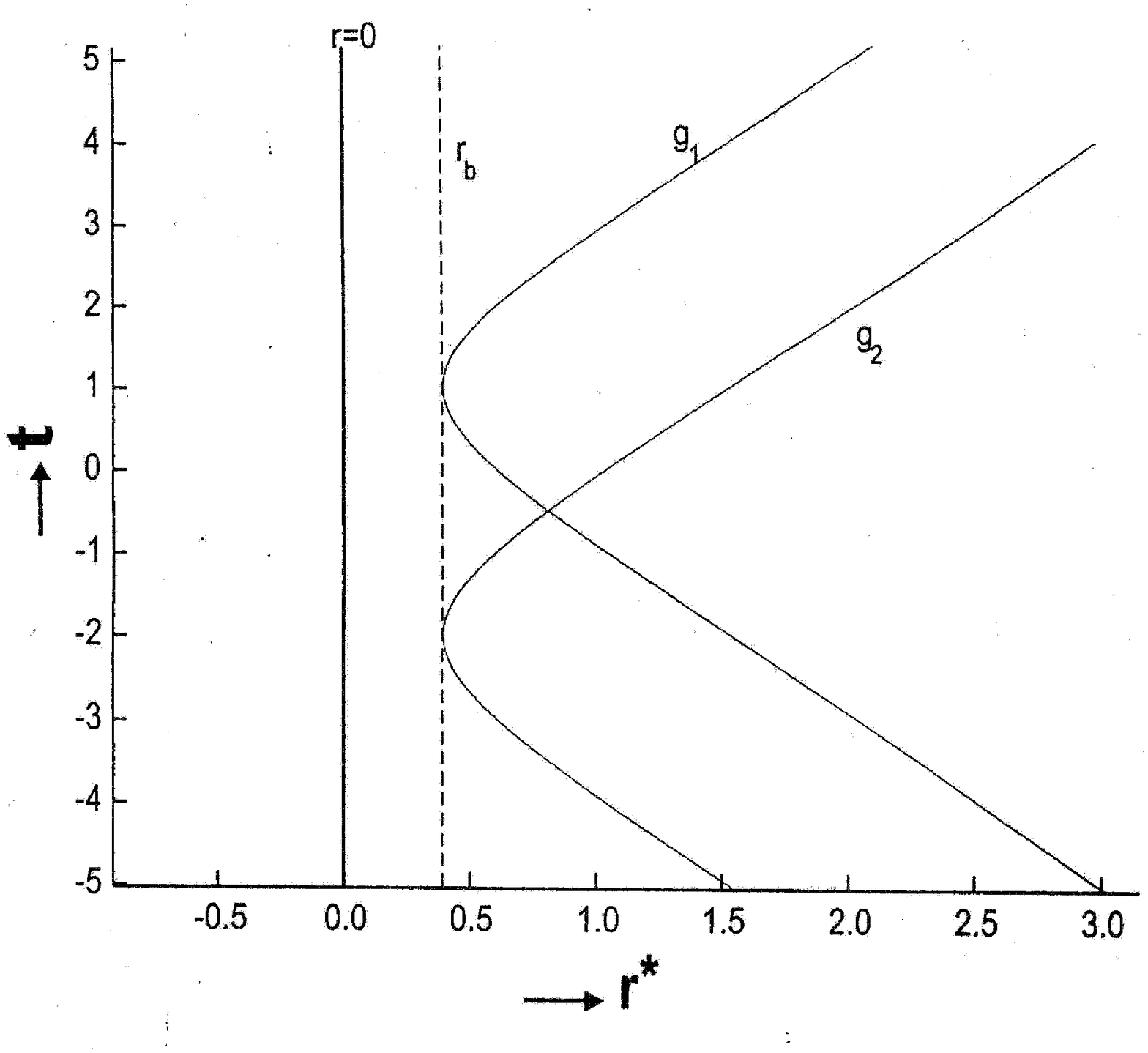, width=.4\textwidth}
{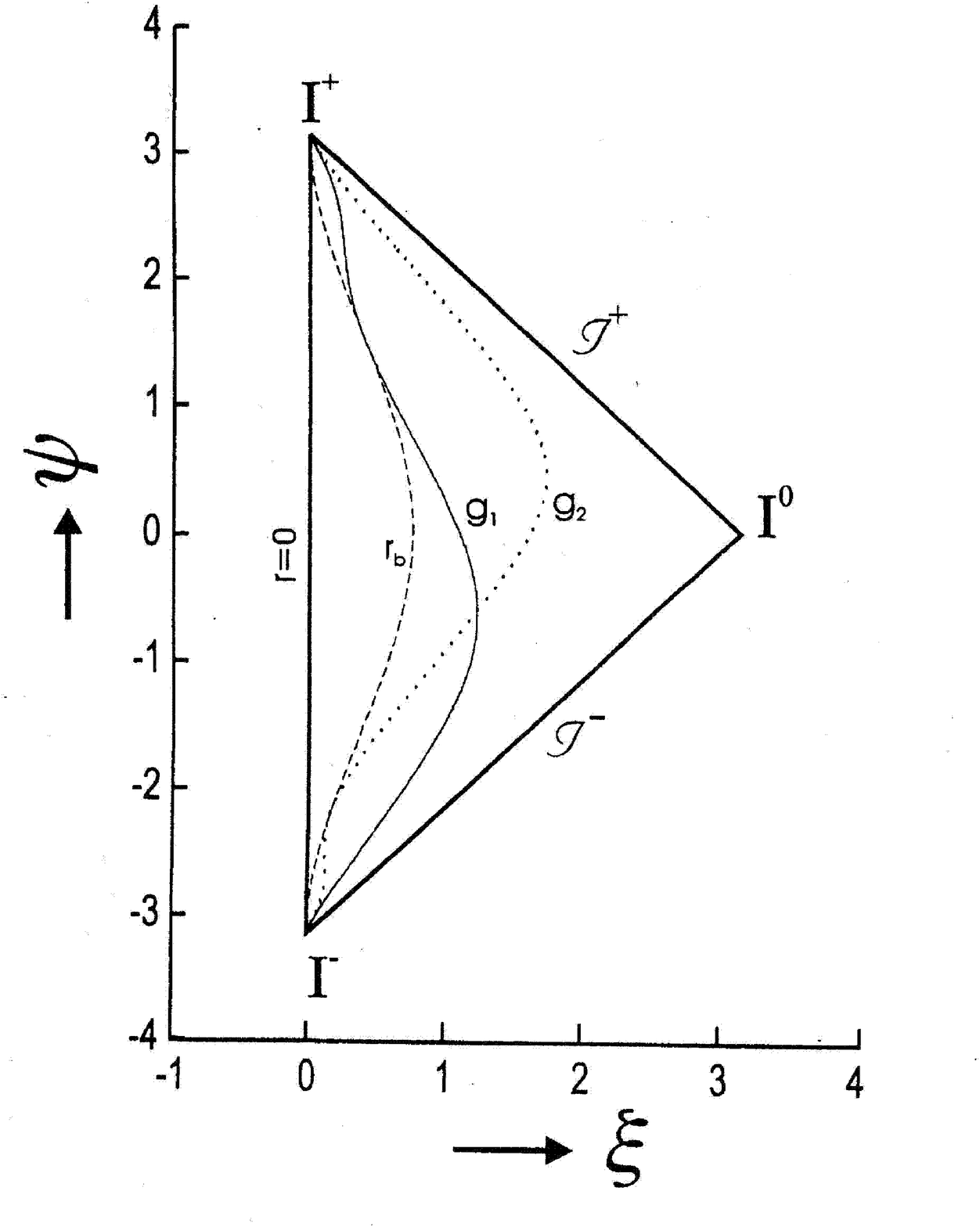, width=.4\textwidth}{Two typical geodesics, $g_{1}$ and $g_{2}$, in
$(t,r^{\ast })$ coordinates, for $k=1$. Note that they all have
the same behaviour, coming in, hitting $r=r_{b}$ (shown as a
dashed line) and going back. Also note that they are parallel. The
only difference between them is the time at which they hit the
classical electromagnetic radius, $t=1$, and $t=-2$. (For
definiteness we have taken $Q=2m$.)}{The same $k=1$ geodesics, as in Figure 1, in the Carter-Penrose
diagram. The behaviour at infinity is made explicit here. They all
come from $I^{-}$\ going along $\vartheta ^{-}$, touch $r=r_{b}$
(dashed line) and proceed
towards $I^{+}$\ along $\vartheta ^{+}$. The difference of the values of $
\psi $ and $\xi $ at which they touch the barrier: $\psi $
$=1.49$, $\xi =0.41$ for $g_{1}$ and $\psi =-2.19$, $\xi $ $=0.16$
for $g_{2}$ (dotted line); changes their appearance in this
diagram.}
Notice that this would be the speed of light, if it took the value
$1$. It takes its extremal values at $r=2r_{b}$, namely $\pm
m/Q$, which are
necessarily less than magnitude $1$. Figure 1 shows the geodesics in a plot of $
t$ against $r^{\ast }$. It is clear that the geodesics coming in
from infinity turn at $r=r_{b}$ and go back to infinity. Thus, no
timelike geodesics from infinity can enter the region $r<r_{b}$.
That region is protected from view by its repulsive nature!

These geodesics can be displayed most meaningfully in a
Carter-Penrose diagram \cite{Hawking}. We consider only the case $Q>m$. In this
case there is only one coordinate patch required, and we only need
to change coordinates to compactify them. For this purpose we
define,
\begin{equation}
\left.
\begin{array}{c}
\psi =\tan ^{-1}\left\{ \left( t+r^{\ast }\right) /R\right\} +\tan
^{-1}\left\{ \left( t-r^{\ast }\right) /R\right\} , \\
\xi =\tan ^{-1}\left\{ \left( t+r^{\ast }\right) /R\right\} -\tan
^{-1}\left\{ \left( t-r^{\ast }\right) /R\right\} ,
\end{array}
\right\}
\end{equation}
where $R$ is a constant with dimensions of length. Thus $r=0$
corresponds to $\xi =0$; $r=\infty $ at $t=0$ to $\xi =\pi $;
$t=-\infty $ to $\psi =-\pi $; $t=\infty $ to $\psi =\pi $.
Hence the Carter-Penrose diagram representing the naked
Reissner-Nordstrom singularity is an isosceles right-angled
triangle standing on its vertex, with the vertical line
representing the singularity. Geodesics start at $I^{-}$, graze
the boundary, $r=r_{b}$, and end at $I^{+}$. Different geodesics
correspond to different values of $\psi $ and $\xi $ at which the
geodesic grazes the boundary. Some typical geodesics and the
boundary, $r=r_{b}$, are displayed in Figure 2 (taking $Q=2m$).

\DOUBLEFIGURE[t]{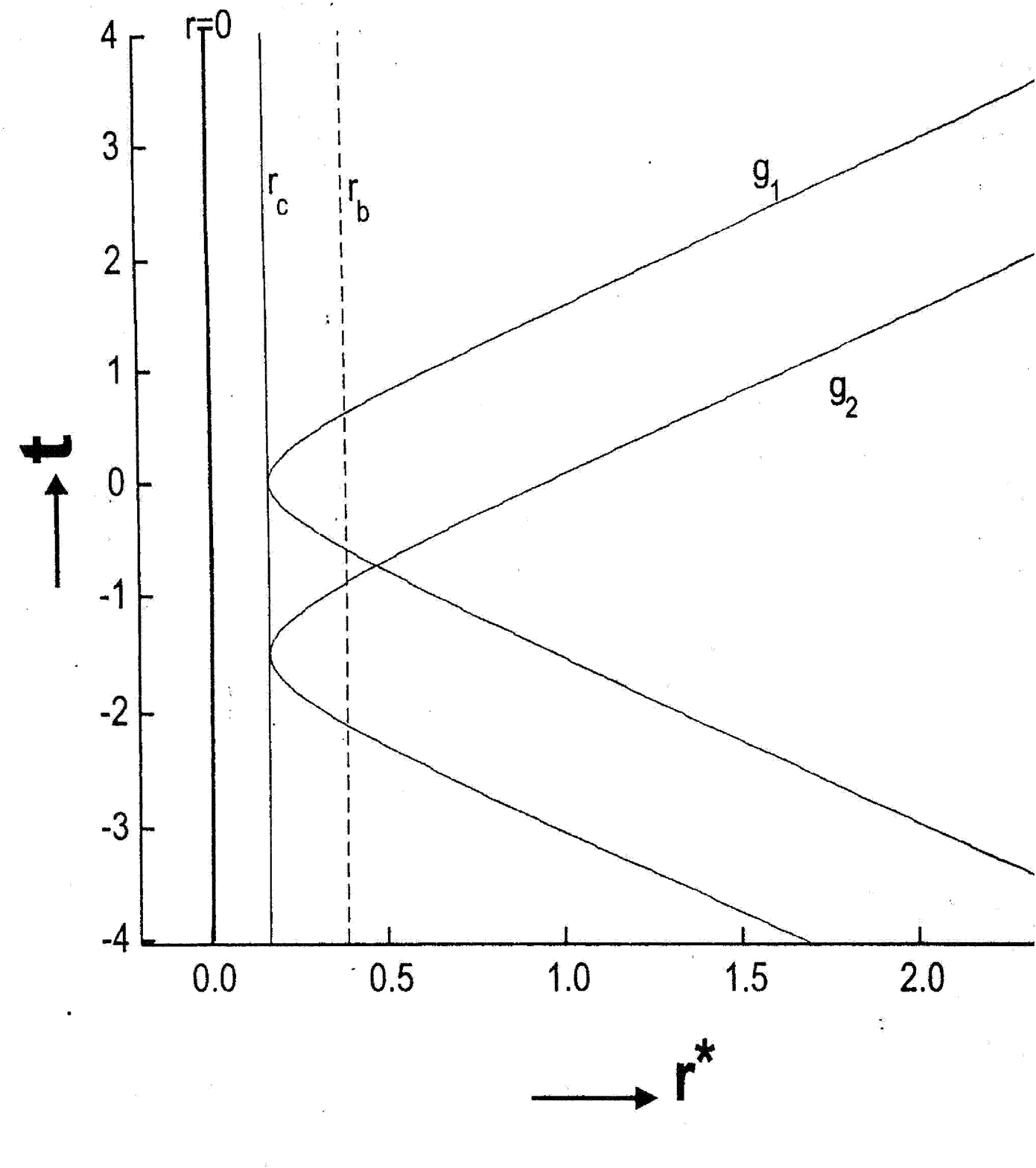, width=.4\textwidth}
{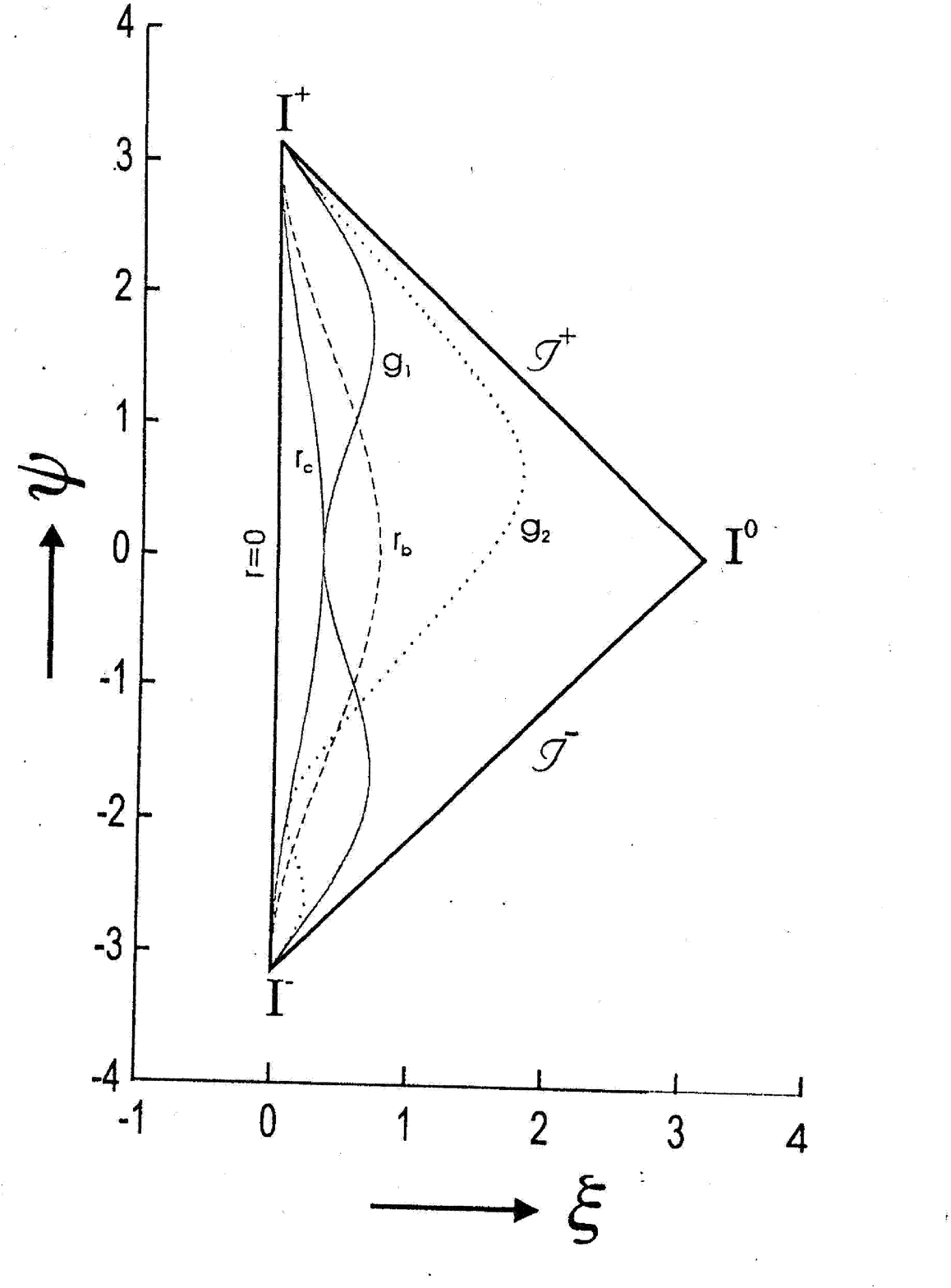, width=.4\textwidth}{Two geodesics in the $(t,r^{\ast })$ coordinates for
$k^{2}=1.5$, touching the new barrier, $r_{c}$, at $t=0$ $(g_{1})$
and $t=-1.5$ $(g_{2})$. The geodesics are again parallel. Notice that the barrier has
gone inwards to $r=r_{c}$ compared with $r=r_{b}$ (dashed line).}{The previous $k^{2}=1.5$
geodesics are shown in the Carter-Penrose diagram. Here $g_{1}$ touches $r_{c}$ at $\psi =0$, $\xi =0.34$ and $g_{2}$ (dotted line) at $\psi =-1.96$, $\xi =0.1$. The geodesics, again, start at $
I^{-}$, go along $\vartheta ^{-}$, touch $r=r_{c}$ and proceed towards $
I^{+} $ along $\vartheta ^{+}$.}

For $k>1$, we get geodesics corresponding to an observer with a
positive velocity at infinity. Putting $k^{2}=1+\varepsilon $, we
find that the geodesics will not turn back at $r=r_{b}$, but
rather at $r_{c}=\left[ -m\pm \sqrt{m^{2}+\varepsilon
Q^{2}}\right] /\varepsilon $. It is easily verified that only the
+ve root is valid. Again, $dr^{\ast }/dt$ takes its maximum value
at $r=2r_{b}$, but now the extremal values are

\[
\left(\frac{dr^{\ast }}{dt}\right)_{max}=\pm \sqrt{\frac{\varepsilon +m^{2}/Q^{2}}{1+\varepsilon}}.
\]
Clearly, for small $\varepsilon $ this tends to the previous
value, but for large $\varepsilon $ it tends to $1$. Now the
boundary moves \textit{back} from $r_{b}$ to $r_{c}\approx
r_{b}-\varepsilon Q^{4}/8m^{3}$, see Figure 3, and the ``clothed"
singularity appears \textit{smaller} to a faster moving observer.
Note that here $k$ is bounded from below by 1, but is not bounded
from above. In the limit as $k$ goes to infinity, $r_{c}$ goes to
zero. The geodesics have the same general behaviour as in the
previous case. Figure 4 shows some typical geodesics in the Carter-Penrose
diagram for $k>1$.

For $k<1$ we put $k^{2}=1-\varepsilon $. The possible reversals are at $
r_{\pm }=\left[ m\pm \sqrt{m^{2}-\varepsilon Q^{2}}\right]
/\varepsilon $. In this case the boundary will move forward from
$r_{b}$ to $r_{-}\approx
r_{b}+\varepsilon Q^{4}/8m^{3}$ while the limit at infinity moves back to $
r_{+}\approx 2m/\varepsilon $, see Figure 5, and the ``clothed"
singularity appears \textit{larger}. These geodesics again start
at $I^{-}$ going along $r=r_{+}$ and then grazing $r=r_{-}$ and
going on to $I^{+}$ along $r_{+}$. Again there can be different
choices of $\psi $ where the geodesic grazes the inner boundary.
Clearly, $k$ is bounded from below by the requirement
that $(m^{2}-\varepsilon Q^{2})$ be +ve. Thus, for a given $m$ and $Q$, $
\varepsilon \leq m^{2}/Q^{2}$. At $\varepsilon =m^{2}/Q^{2}$ we get $
r_{+}=r_{-}=2r_{b}$. This gives the geodesic as $r=2r_{b}$. In the
other limit, as $k$ tends to 1 we see that the outer limit,
$r_{+}$, tends to infinity. In general, the geodesics are bounded by the two boundaries$
r=r_{+}$ and $r=r_{-}$, being asymptotic to the former and
tangential to the latter. Some typical geodesics are displayed in
a Carter-Penrose diagram, in Figure 6.

\DOUBLEFIGURE[t]{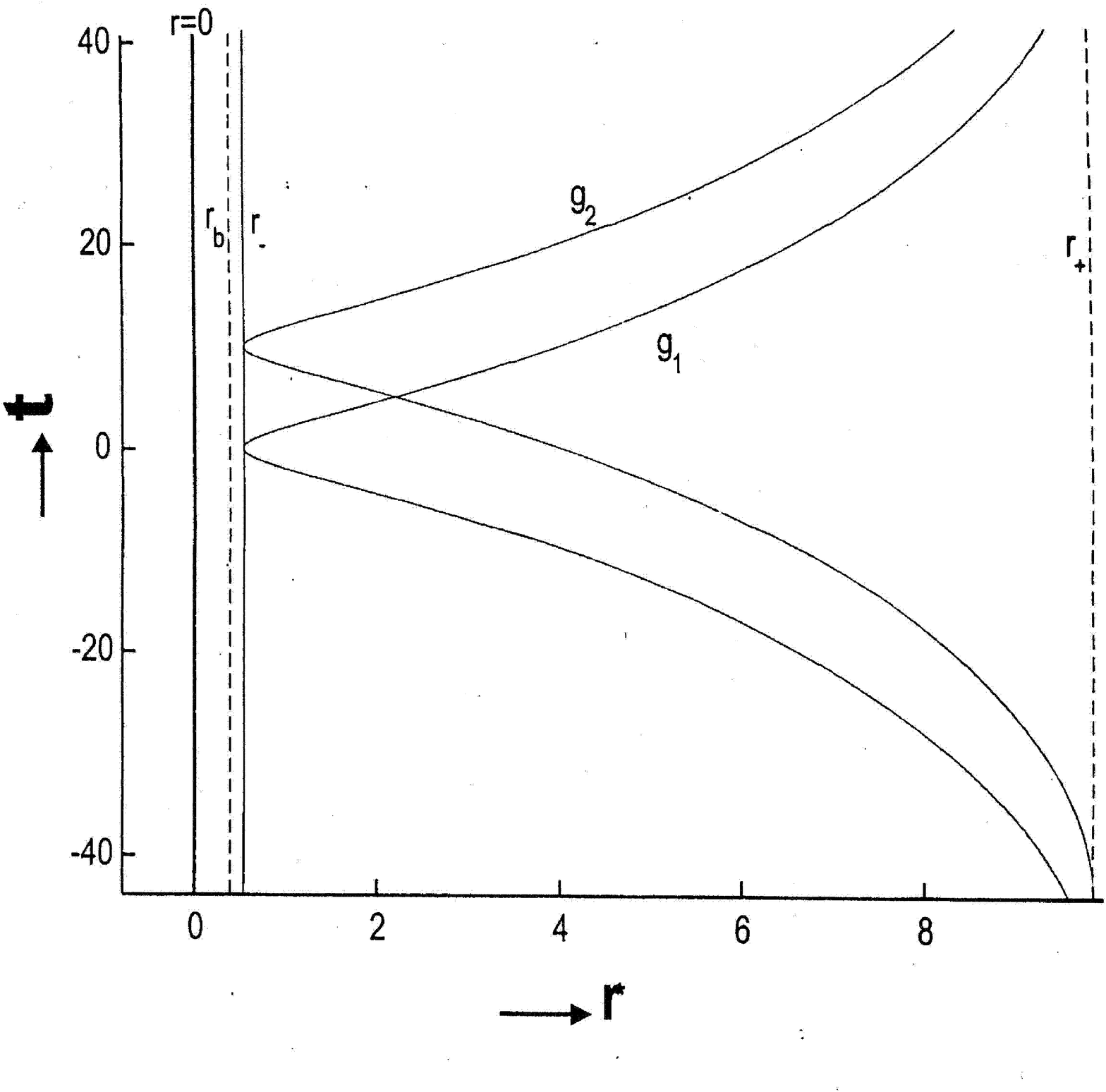, width=.4\textwidth}
{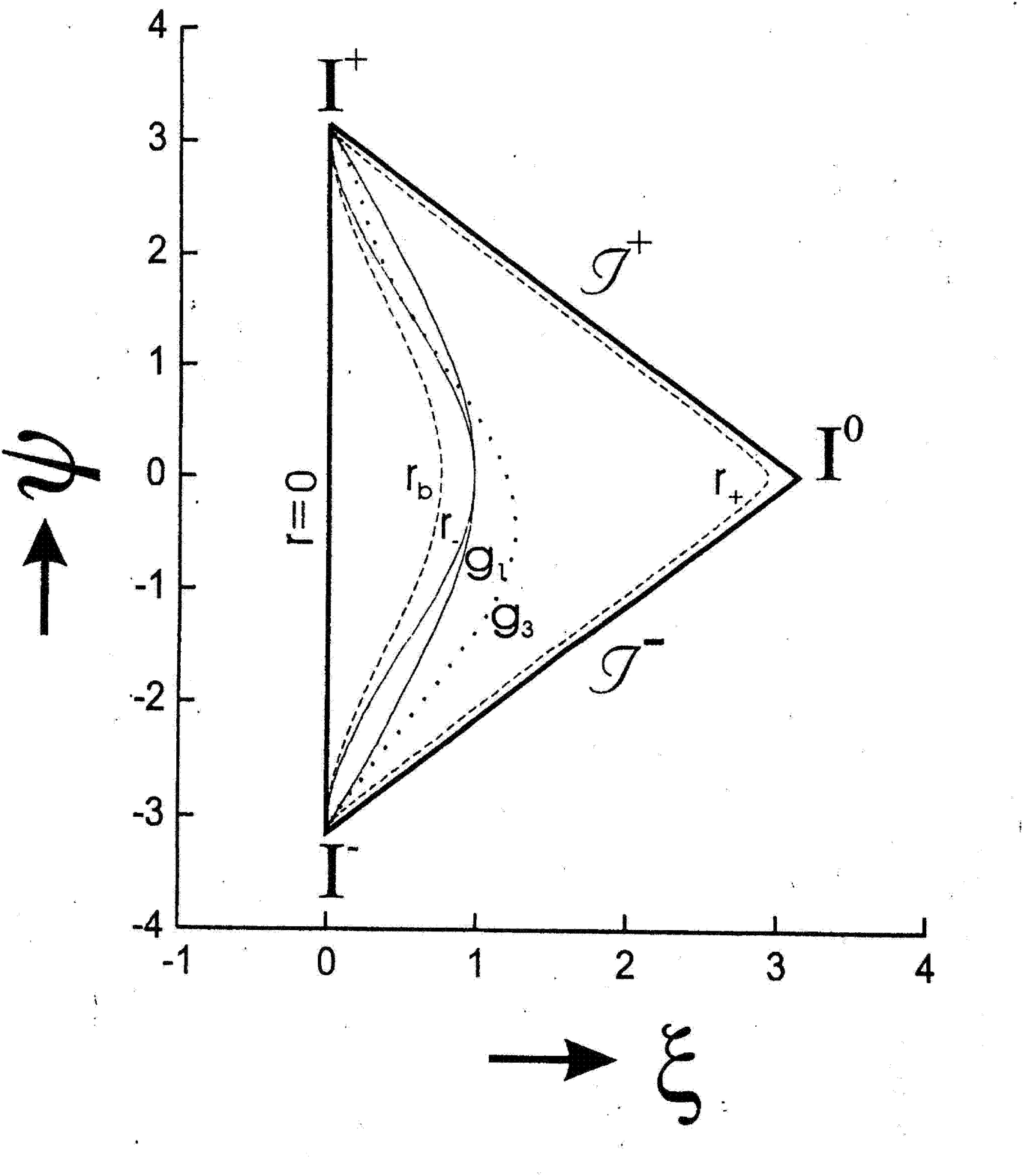, width=.4\textwidth}{Two geodesics in the $(t,r^{\ast })$ coordinates for
$k^{2}=0.9$. Now the geodesics start at $r=r_{+}$ (dashed line) in
the infinite past and go in to $r=r_{-}$ before going back out to
$r=r_{+}$. We have taken one geodesic, $g_{1}$, touching the inner boundary
at $t=0$, and the other, $ g_{2}$, at $t=10$. Notice that the inner boundary
lies outside the classical electromagnetic radius.}{Two geodesics for $k^{2}=0.9$ in the Carter-Penrose
diagram. Here $g_{1}$ is the same as before (touching $r_{-}$ at
$\psi =0$, $\xi =0.98 $) but we have chosen a new geodesic, $g_{3}$,
instead of $g_{2}$ (touching $ r_{-}$ at $\psi =1.43$, $\xi $ $=0.55$),
to be able to show the relevant features in both diagrams. The latter geodesic would
touch $r_{-}$ at $t=1$ in the previous diagram. The geodesics
still start at $I^{-}$, but now go along $r=r_{+}$, touch
$r=r_{-}$ and go on to $I^{+}$\ along $r=r_{+}$.}

We need to interpret the significance of $k$ or $\varepsilon$.
When $k=1$, we have the energy at infinity equal to the rest
energy. Thus these geodesics correspond to zero kinetic energy.
For $k>1$ there is extra energy of motion at infinity and hence
these geodesics correspond to faster moving observers. However,
for $k<1$ the energy of motion is less than zero! Hence
these geodesics must correspond to negative energy. In fact, as $
r\rightarrow \infty $, $\stackrel{.}{t}\rightarrow k$. Also,
$\stackrel{.}{t} $ corresponds to the ratio of total energy to
rest energy. Thus $\varepsilon =k^{2}-1$ corresponds to
(KE/RE)(2+KE/RE), where KE is the kinetic energy
and RE the rest energy, at infinity. In the high energy limit, then, $
\varepsilon $ corresponds to (KE/RE)$^{2}$, while in the low
energy limit it
corresponds to 2(KE/RE). In the intermediate energy it comes out approximately 3(KE/RE). Now $k<1$ corresponds to negative $\varepsilon$, and hence negative
energy as defined at infinity. It must be borne in mind that these
geodesics are never at infinity, but remain trapped near the
singularity.

\section{The Electron as a Naked Singularity}

In gravitational units $(c=G=1)$, $Q=1.4\times10^{-34}$cm and
$m=6.8\times10^{-59}$cm for the electron. Thus, if we were to take the
general relativistic (GR) description of the electron seriously,
we must consider it as a naked singularity. Of course, we should
really treat the electron as a charged Kerr (or Kerr-Newmann, K-N)
singularity and not a Reissner-Nordstrom singularity. However, for
the present purposes, the essential features are already
highlighted by the simpler analysis.

We see that here $r_{b}$ is the classical electromagnetic radius,
$1.4\times10^{-13}$! This is the size the electron would appear to be,
to an observer at rest at infinity. However, a faster moving
observer would see it shrunk arbitrarily smaller to $r_{c}$. In the
high-energy limit, the size would decrease to $Q/\sqrt{\varepsilon}$, thus decreasing
inversely as the kinetic energy. This is strongly reminiscent of
QT!

One is used to non-local effects in GR, where the global aspect of
the theory plays a fundamental role. Here, there will be non-local
effects associated with the electron corresponding to the negative
energy paths. In particular, an electron passing through a single
slit will behave differently from one passing through a double
slit, because of the negative energy paths that may be blocked or
pass through the second slit. Once again, strongly reminiscent of
the QT! Remember that the further out the negative energy paths
go, the less the energy associated with them. As such, they would
be disturbed more readily. Hence, the non-local effects of the
electron would be more difficult to maintain further out.

Since the negative energy paths can not reach out to infinity,
they can not be seen by outside observers. What, then, is their
significance? They may still be physically relevant when dealing
with the interaction of two electrons. We have no means available
of getting an exact solution for the two electrons. It seems
reasonable to conjecture, however, that the approximate solution
would allow the two electrons to interact through their negative
energy paths as well as their positive energy paths. This would
provide a correction to the usual calculation of the interaction.
This is strongly reminiscent of the renormalisation calculation
involving virtual particles!

It had been demonstrated earlier \cite{Qadir} that for the Kerr metric the
total angular momentum and the axial component are well defined,
but the other two components are not. This, too, is strongly
reminiscent of the QT!

\section{Problems --- Incorporation of Angular Momentum and Dealing with
Protons}

It is clear that we can not neglect the spin angular momentum of
the electron. The spin angular momentum per unit mass of the electron is, $
a=1.9\times10^{-11}$cm. Thus, if the $a$ enters into the calculations
like $Q$, the scale will be literally astronomical. As we have no
analysis for the naked K-N singularity so far, we are unable to
say whether, or not, the very high $a$ of the electron would create
a problem for us. It seems likely to us that the dragging of
inertial frames will change the way a enters into the calculations
for the geodesics, as compared with $Q$.

There is another worry. While discussing various aspects of the
behaviour of electrons treated as naked singularities we have
remarked on its similarity with that in Quantum Theory. However,
there is no way that Planck's constant can enter into the
analysis. Of course, if we incorporate the spin angular momentum
Planck's constant will automatically enter, but will not be
provided by this analysis.

If electrons can be treated as naked singularities, why not
protons? If they could be so treated, they should have a size of
$7\times10^{-16}$cm. This does
not seem reasonable. Since the proton already shows structure at a scale of $
10^{-13}$cm, it can not be thought of as a point particle. In
fact, it is best described as a bound state of three quarks, which
could perhaps be thought of as three point particles. It would
not, then, be described by one, but three, naked singularities
close together. Further, these singularities would not be
Reissner-Nordstrom singularities, but Einstein-Yang-Mills
singularities. Clearly, there can be no exact description of this
situation. Though there is a solution available \cite{Bartnik} to the
Einstein-Yang-Mills equations for a single source, it is not
exact. (In fact it does not even have a singularity, though it is
a solution for a point source.) As such, there is no reason to
suppose that the existence of the proton poses a problem for our
argument treating the electron as a naked singularity. It must be
admitted, however, that there is much work required to verify
whether this proposal would be workable in a more general context,
not only including the K-N, but other naked singularities and
finding methods for dealing with approximate solutions rigorously.

\section*{Acknowledgments}
One of us (AQ) is most grateful to David Finkelstein for very
valuable criticism of the manuscript and Prof. Remo Ruffini for
hospatility at ICRA, where he got the opportunity to meet Prof.
Finkelstein.

\end{document}